\documentclass[prc,aps,amsmath,amssymb,nofootinbib,superscriptaddress,12pt]{revtex4-2}
\usepackage{epsfig}
\usepackage[bookmarksnumbered,bookmarksopen,colorlinks,citecolor=blue,linkcolor=blue]{hyperref}
\usepackage{amsmath}
\usepackage{amssymb}
\usepackage{tipa}
\usepackage{float}

\begin{document}

    \title{Production probability of super-heavy nuclei in fusion }

\author{Ning Wang}
\email{wangning@gxnu.edu.cn}\affiliation{ Department of Physics,
	Guangxi Normal University, Guilin 541004, People's Republic of
	China }
\affiliation{ Guangxi Key Laboratory of Nuclear Physics and Technology, Guilin 541004, People's Republic of
	China }
 
\author{Cheng Li}
\affiliation{ Department of Physics,
	Guangxi Normal University, Guilin 541004, People's Republic of
	China }
\affiliation{ Guangxi Key Laboratory of Nuclear Physics and Technology, Guilin 541004, People's Republic of
	China }

\author{Hong Yao}
\affiliation{ Department of Physics,
	Guangxi Normal University, Guilin 541004, People's Republic of
	China }
\affiliation{ Guangxi Key Laboratory of Nuclear Physics and Technology, Guilin 541004, People's Republic of
	China }
			
    \begin{abstract}
   The synthesis of super-heavy nuclei (SHN) through fusion reactions is a critical area of nuclear physics, offering insights into nuclear stability and the limits of the periodic table. However, theoretical predictions of evaporation residue cross sections $ \sigma_{\rm {ER} }$ remain challenging due to large uncertainties arising from complex reaction mechanisms and sensitive model parameters. In this work, the average production probabilities of SHN with atomic number $Z\ge 110$ are systematically analyzed, based on barrier tunneling concept. Together with the empirical barrier distribution method for describing capture, a phenomenological formula, EBD3,  is proposed, which reproduces 58 measured $ \sigma_{\rm {ER} }$ within one order of magnitude, with a root-mean-square deviation of 0.369.  The formula successfully captures key quantities in fission-like process, including fission barrier height, mass asymmetry, depth of capture pocket and the effective fusion barrier height.  Predictions for the synthesis of element 119 are presented, identifying promising projectile-target combinations such as 
$^{50}$Ti + $^{249}$Bk with a maximum cross section of $54.9_{-29.0}^{+61.3}$ fb at excitation energy of 36.5 MeV. The maximum cross section falls to $3.2_{-1.7}^{+3.6}$ fb  for $^{54}$Cr + $^{243}$Am  at the optimal incident energy of 244 MeV.

    \end{abstract}

     \maketitle

\newpage
  
   \begin{center}
  	\textbf{ I. INTRODUCTION }\\
  \end{center}

Synthesis of super-heavy nuclei (SHN) through fusion reactions is a field of intense studies in nuclear physics \cite{Hoff98,Hoff02,Mori04,Mori09,Ogan15,Ogan15a,Ogan17,Khu20,Tana22,Ogan24,Ogan25,Ogan26,Gan26,Itkis22,Adam04,Zag08,Wang11,Wangnan12,ZhuL14,Adam20,Guolu23,Nasirov24,Pei24,WangB25,LiJJ25,ZhangFS,ZhangHF26,ZhangYH26}. Synthesizing SHN and studying their properties (such as binding energies, decay modes, and half-lives) provides a stringent test for advanced nuclear models, including the microscopic shell model \cite{Bend99,Sam05,Pei16} and relativistic mean-field theories \cite{Lala,Abu12,Lu14,Zhou16}. The synthesis of new elements physically extends the periodic table and the understanding of the limits of nuclear stability, with the current frontier at element 118 (Oganesson) and efforts underway to create elements 119 and 120 \cite{Khu20,Tana22,Wang11,Wangnan12,ZhuL14,Adam20,Nasirov24,WangB25,LiJJ25,ZhangFS,ZhangHF26}. At these extremes, the high nuclear charge accelerates inner electrons to relativistic speeds. Furthermore, the properties of superheavy nuclei, such as their fission barriers, half-lives, and decay modes, are crucial input parameters for models that simulate the r-process \cite{Zhao19} and aim to explain the cosmic abundance of elements heavier than iron.

In heavy-ion fusion reactions leading to the synthesis of SHN, the evaporation residue cross section is usually expressed as a product of three parts: 	$\sigma_{\rm ER} = \sigma_{\rm cap} \times P_{\rm CN} \times W_{\rm sur} $ \cite{Wang11}. Here, $\sigma_{\rm cap}$, $P_{\rm CN}$ and $W_{\rm sur}$ denote the capture cross section of the colliding nuclei overcoming the entrance-channel Coulomb barrier, the probability of the compound nucleus formation after the capture and the survival probability of the excited compound nucleus, respectively. It is thought that the current uncertainty in calculated $\sigma_{\rm ER}$ is at least 1–2 orders of magnitude \cite{Nasirov11,Loveland,Loveland15}. The large uncertainties in calculating $\sigma_{\rm ER}$ arise because one attempts to use a simplified theoretical framework to describe a reaction that is a chain of multiple, highly complex, and interconnected physical processes. The models for each step must make approximations to tackle complex quantum many-body problems (such as nucleon-nucleon interactions, collective excitation, and dissipative effects), while also being heavily dependent on nuclear structure parameters (like deformation \cite{Guolu23} and fission barriers \cite{Moll15,Wang24}) that cannot be determined precisely. Minor deviations from these approximations and parameters are propagated and amplified step-by-step along the reaction chain, ultimately leading to significant discrepancies in theoretical predictions.

For example, $P_{\rm CN}$ describes whether, after two nuclei come into contact, they fuse to form a compound nucleus or separate via quasi-fission which depends on various factors in the entrance channel, such as the mass asymmetry of the two nuclei, their relative orientation, and angular momentum. Different theoretical models (e.g., the di-nuclear system model \cite{Adam20,Nasirov24}, transport models \cite{Guolu23,Li25,Li20}) handle these complex factors differently, leading to large uncertainties of $P_{\rm CN}$ \cite{Loveland15}. More critically, $P_{\rm CN}$ itself is difficult to measure directly and must be extracted from experimental data. However, in experiments, the mass or angular distributions of quasi-fission products often overlap with those from deep inelastic collisions and fusion-fission, making clear separation challenging \cite{Itkis22,Yao24}. In the calculations of the survival probability of the compound nucleus based on statistical models \cite{Reis85,EBD}, $W_{\rm sur}$ strongly depends on the fission barrier height and its temperature-angular-momentum dependence, in addition to the level density parameters, nuclear masses, nuclear viscosity and other input parameters. An uncertainty of 1 MeV in the fission barrier height can lead to a difference of about one order of magnitude in the survival probability for the 4n evaporation channel \cite{Nasirov11,Loveland,Loveland15}. Furthermore, the production cross sections of SHN with charge number $Z\ge 110$ are typically smaller than tens of picobarn, which results in that the number of available experimental data of $\sigma_{\rm ER}$ is limited and not large enough to well constrain all of the model parameters in $\sigma_{\rm cap}$, $P_{\rm CN}$ and $W_{\rm sur}$.

It is therefore necessary to propose a new model or phenomenological method to systematically analyze the measured evaporation residue cross sections and give reliable predictions for the synthesis of new SHN. In nuclear physics, some phenomenological approaches (e.g., the Bass potential \cite{Bass74,Bass80}, the Siwek-Wilczyńska barrier distribution method \cite{SW04}, etc.) are also commonly used for guiding experiments, in addition to microscopic models.   In \cite{Wang26}, an analytical formula  (EBD2.2)  based on the empirical barrier distribution (EBD) method \cite{SW04} is proposed for a systematic description of the capture cross sections at near-barrier energies from light to superheavy reaction systems. The uncertainties of EBD2.2 are within half an order of magnitude at sub-barrier energies and approximately $21.6\%$ at energies well above the Coulomb barrier for a systematic description of 426 datasets of measured capture excitation functions \cite{Wang26}. In this work, the systematic behavior of the production probabilities of SHN will be further investigated based on the barrier tunneling concept and a new version (EBD3) of the formula will be proposed for calculating not only $\sigma_{\rm cap}$ but also $\sigma_{\rm ER}$ of super-heavy fusion systems. Our goal is not to analyze the fusion mechanism from microscopic or macroscopic dynamics calculations, but rather to provide a tool with simple computation, and the ability to systematically describe experimental data, under circumstances where the uncertainties of existing theories are large (1–2 orders of magnitude), so as to guide experimental design.

 \begin{center}
	\textbf{ II. PRODUCTION PROBABILITY OF SUPER-HEAVY NUCLEI }\\
\end{center}
  
 In this work, we focus on the fusion reactions leading to the synthesis of SHN with $Z\ge 110$. The average production probability of SHN is defined as $P=\langle P_{\rm CN} \times W_{\rm sur} \rangle$ and the total evaporation residue cross section is written as,
 \begin{equation}
 \sigma_{\rm {ER} }(E_{\rm c.m.})  =\sigma_{\rm {cap} }(E_{\rm c.m.})\times P(E_{\rm c.m.}).
 \end{equation}
The capture cross section $\sigma_{\rm {cap} }$ is calculated with EBD2.2 \cite{Wang26}, 
\begin{equation}
	\sigma_{\rm {cap} } (E_{\rm c.m.})=\pi R_B^{2}  \frac{W}{\sqrt{2}E_{\rm c.m.}}
	[X  {\rm erfc}(-X)+\frac{1}{\sqrt{\pi}}\exp(-X^{2})].
\end{equation}
Where $X =\frac{E_{\rm c.m.}-V_B}{\sqrt{2}W} $.  $E_{\rm c.m.}$ denotes the center-of-mass incident energy. $V_B$ and $W$ denote the centroid and the standard deviation of the Gaussian function, respectively.  $R_B$ denotes the barrier radius. The parameters of EBD2.2 are fixed by the global fit to 426 measured capture excitation functions and are not adjusted in the present work.

The average production probability is written as, 
 \begin{equation}
	P(E_{\rm c.m.}) = s P_{\rm mac}(Z,\eta)  P_1(E_{\rm c.m.}) P_2(E_{\rm c.m.}).
\end{equation}
$s$ is a scale factor introduced for SHN to consider the relative height of the fission barrier. $P_{\rm mac}=(1-P_0)^2$ denotes the macroscopic part of the  survival probability of the di-nuclear system (DNS), with a smooth fission-like (i.e., quasi-fission and fusion-fission) probability of $P_0$. Because fusion is a chain of multiple, highly complex, and interconnected physical processes, the survival probability of DNS is propagated and shrinks step-by-step along the reaction chain. The square reflects the fact that the DNS must survive two distinct stages of the fission-like process: a fast dynamical stage during which the system
evolves toward a compact shape, and a subsequent slow quasi-stationary stage near the saddle point \cite{Nasirov24,Pal24,Alb20}. $P_1$ and $P_2$ denote the energy-dependent terms for the survival probability of DNS and that of the compound nucleus, respectively. The traditional calculations of the production probability depend heavily on the reaction's angular momentum. In this work, we focus on the average production probability. 

From Fig. 1(b) in \cite{Wang11}, we note that the quasi-fission barrier heights are related to the charge number $Z$ of the compound nucleus and the mass asymmetry $\eta=|A_1-A_2|/(A_1+A_2)$. They systematically decrease with increasing $Z$ and decreasing $\eta$. Considering that the macroscopic fission-like probability is closely related to the barrier heights of quasi-fission and fusion-fission, it should systematically increase from a value of zero for light-asymmetric fusion systems to about one for heavy-symmetric reactions such as Sn+Sn. $P_0$ is therefore described by a Fermi function,
 \begin{equation}
	P_0 =\frac{1}{1+e^{-(Z-Z_0)/a}}.
\end{equation}
The borders between fusion-dominant systems after capture and fission-dominant ones are determined by the  parameters $Z_0=72 (1+\eta^4)$ and $a=3 (1+\eta^4)$. The power $\eta^4$ is chosen empirically to provide a realistic contour of  $P_{\rm mac}$ consistent with the known systematics of quasi-fission barrier heights \cite{Wang11}.

Based on barrier penetration concept, the energy-dependence of the DNS survival probability is expressed as,
 \begin{equation}
	P_1 =1-\frac{1}{1+e^{-(E_{\rm DNS}^*-U_0)/\hbar \omega}}.
\end{equation}
Here, $E_{\rm DNS}^*$ denote the excitation energy of the DNS. Considering that a number of nucleons are transferred during the fission-like process of DNS, $E_{\rm DNS}^*$ should be larger than the excitation energy at capture position $E_{\rm cap}^*$. In \cite{YaoH}, $E_{\rm cap}^*$ is systematically investigated with the time-dependent Hartree-Fock (TDHF) theory for 144 fusion systems with nearly spherical nuclei. It is found that $E_{\rm cap}^*$ is about $5.2\%$ of the compound-nuclei excitation energy $E^*$. We therefore take $E_{\rm DNS}^*=20\% E^*$ as a working assumption in the
calculations. A direct microscopic test of this assumption using the improved quantum molecular dynamics (ImQMD) model \cite{Li20,Yao24} is provided in Sec. III B (see Fig. 9). $U_0$ denotes the effective tunneling barrier (ETB) \cite{Xie26} for the DNS to break up, with a value of $U_0=0$ for SHN with $Z\ge 110$. $\hbar \omega = 1.2 $ MeV denotes the curvature of the ETB.

Inspired by Bohr-Wheeler decay width in the statistical model \cite{Lest09}, the survival probability of the compound nucleus is written as,
 \begin{equation}
	P_2 =Y e^{-(E^*-B_{\rm f})/E_D}.
\end{equation}
$B_{\rm f}$ denotes the fission barrier height predicted with WS4 \cite{WS4} and $E_D=20$ MeV denotes the damping constant which is close to the default values adopted in the statistical models, such as  HIVAP \cite{Reis85} and KEWPIE2 \cite{EBD}. Here, $Y$ denotes an energy-dependent structure factor for the formation of the compound nucleus. It quantifies the deviation from the macroscopic trend due to the specific entrance-channel properties.
For hot fusion, we set
\begin{equation}
	Y =\frac{1}{1+e^{-2\pi [E_{\rm c.m.}-(V_B+B_{\rm f})]/E_D}},
\end{equation} 
to consider the extra-push effects \cite{Bass74,Swiat82} caused by the deformed actinide targets. $E_{\rm c.m.}$ and $V_B$ denote the center-of-mass incident energy and the capture barrier height \cite{Wang26}, respectively.  
For cold fusion, we set
\begin{equation}
	Y = e^{-[E_{\rm DNS}^*-\frac{1}{2}(B_{\rm f}+B_{\rm cap}) ]/\hbar \omega},
\end{equation}
to effectively consider the stabilizing shell effects associated with the double-magic $^{208}$Pb or $^{209}$Bi targets.  $B_{\rm cap}$ denotes the depth of the capture pocket  in the entrance-channel nucleus-nucleus potential calculated by using the Skyrme energy density functional \cite{liumin,EP}.  

As a parameterized factor, the scale factor $s$ collectively absorbs the neglected effects in realistic reactions, such as angular momentum dependence, dissipation effects, viscosity, etc.  Since the evaporation residue cross sections span several orders of magnitude, using base-10 exponential form allows for a more stable fitting of the experimental data. For hot fusion, the large mass-asymmetry makes the entrance-channel pocket depth $B_{\rm cap}$ a decisive
quantity and the scale factor is empirically set as
 \begin{equation}
	\log_{10} s =    (B_{\rm min}-B_{\rm max}) B_{\rm cap}/(B_{\rm f} {\hbar \omega})   .
\end{equation}
Here, $B_{\rm min}=\min(B_{\rm cap},B_{\rm f})$ denotes the minimal value between $B_{\rm cap}$ and $B_{\rm f}$, and $B_{\rm max}=5.33 $ MeV denotes the maximal fission barrier height of known Flerovium with almost the largest $\sigma_{\rm {ER} }$ in measured hot fusion reactions \cite{Ogan15}.  

For cold fusion, where both collision partners are near-spherical, the shell effect plays a more prominent role in stabilizing the compound
nucleus against fission, and the scale factor is set as
\begin{equation}
	\log_{10} s =  (B_{\rm f}+\Delta-B_{\rm max})/B_{\rm min}  ,
\end{equation}
with the predicted maximal fission barrier height $B_{\rm max}=6.23 $ MeV from WS4 for SHN around the center of "island of stability". $\Delta$ denotes the corresponding shell gap \cite{Mo16} in the compound nucleus, which is to effectively consider the influence of barrier width in cold fusion. 
 For systems with $B_{\rm cap}$ or $B_{\rm f}$ smaller than zero-point energy of $\approx 1.0$ MeV \cite{Rein78},  fusion is excluded in the present model. 
 
\bigskip
\bigskip 
 
\begin{center}
	\textbf{ III. RESULTS AND DISCUSSIONS }\\
\end{center}

\begin{figure}
	\setlength{\abovecaptionskip}{0.5cm}
\includegraphics[angle=0,width=0.5\textwidth]{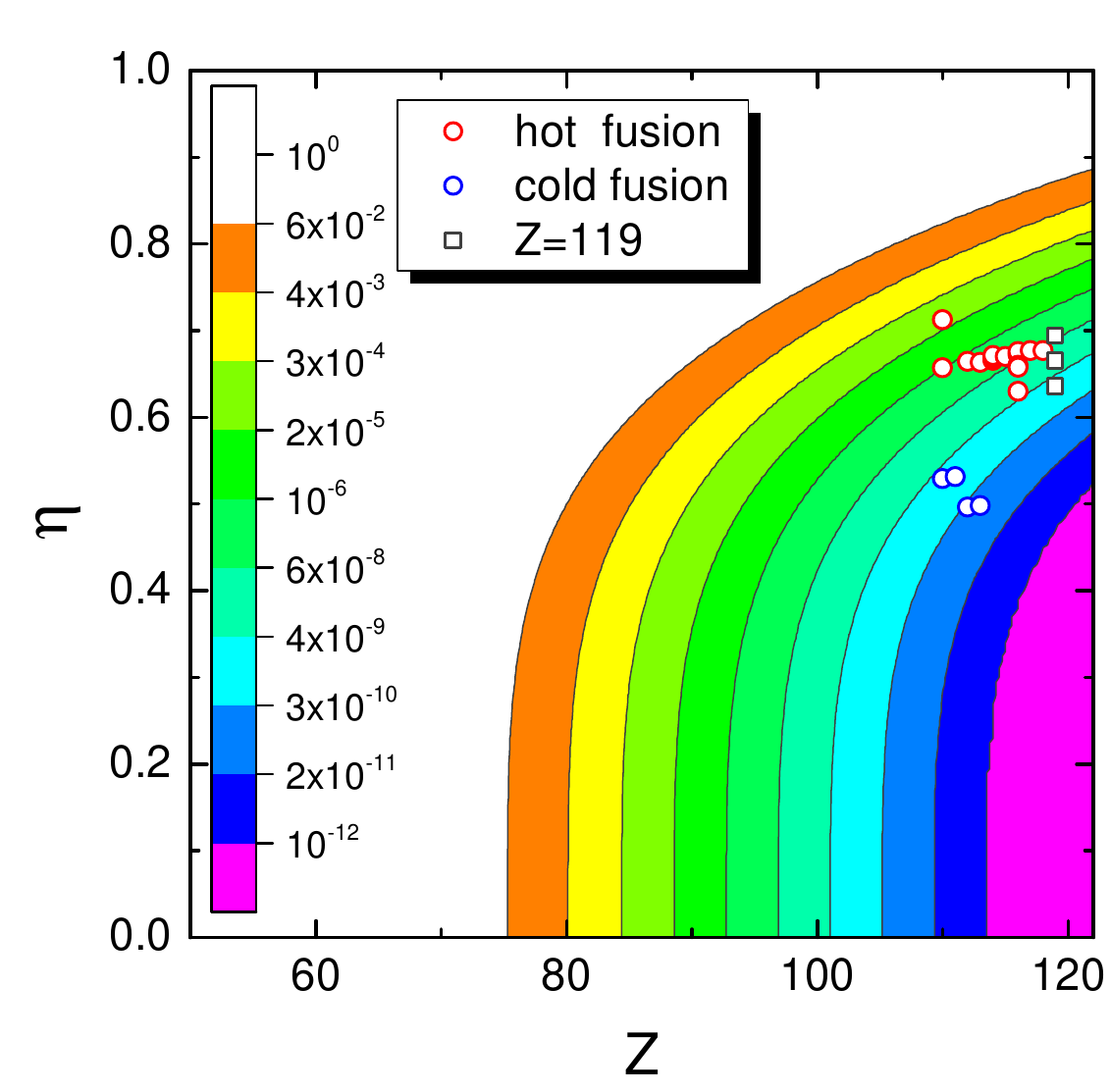}	
	\caption{ Contour plot of the macroscopic part $P_{\rm mac}$ of the DNS survival probability. The red circles and the blue ones denote known hot and cold fusion reactions producing SHN with $Z\ge 110$, respectively. The squares denote the three reactions leading to the synthesis of element 119: $^{54}$Cr+$^{243}$Am, $^{50}$Ti+$^{249}$Bk and $^{45}$Sc+$^{249}$Cf.       }
\end{figure} 

In this section, we firstly present the predicted evaporation residue cross sections for a number of cold and hot fusion reactions. Then, we will discuss the uncertainties of model predictions. In addition, we will also present some results from statistical model and microscopic dynamics simulations for comparison. Finally,  the limitations of EBD3 will be discussed.

\begin{center}
	\textbf{ A. Evaporation residue cross sections }\\
\end{center}

 \begin{figure}
	\setlength{\abovecaptionskip}{0cm}
\includegraphics[angle=0,width=0.5\textwidth]{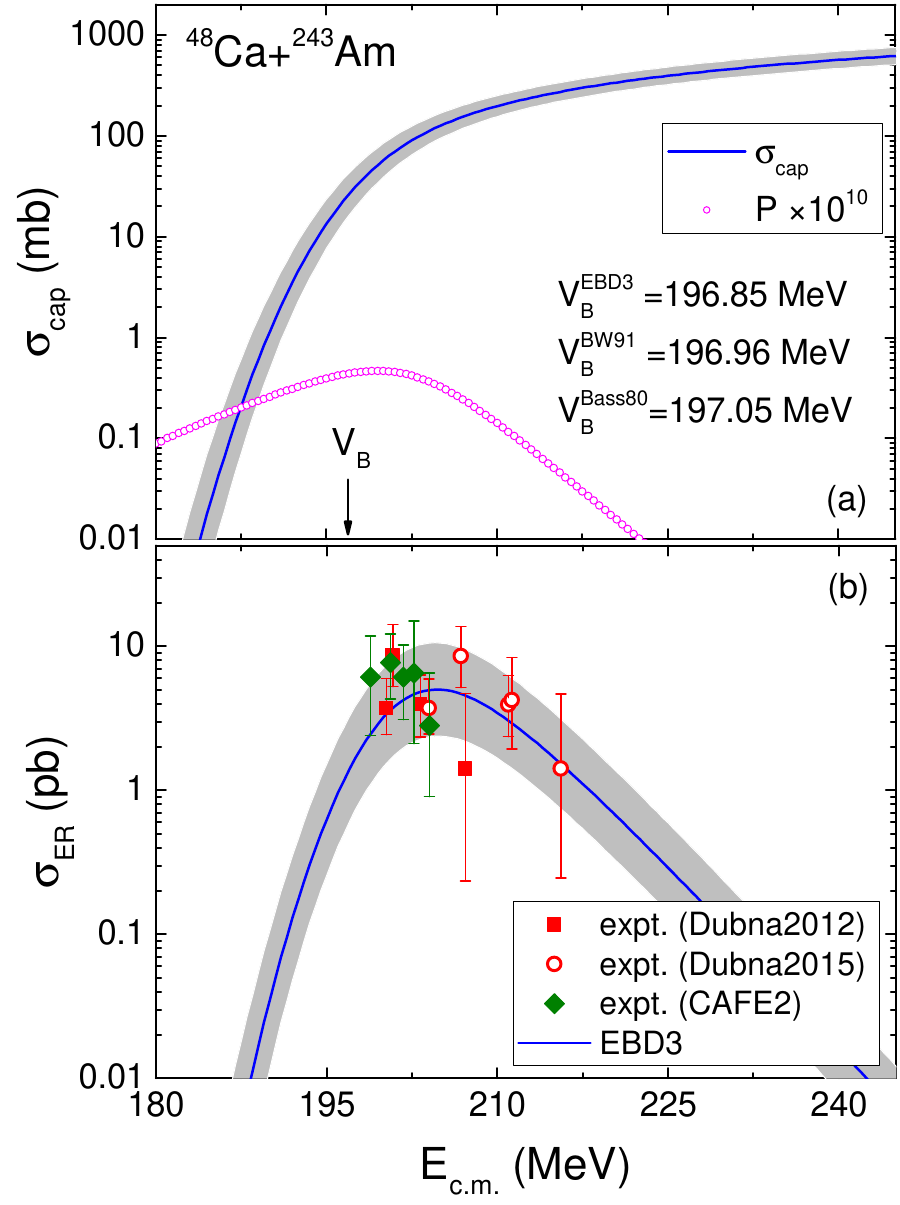}	
 	\caption{ (a) Predicted capture excitation function for  $^{48}$Ca + $^{243}$Am with EBD3. The circles denote the production probability $P$ of SHN calculated by  Eq.(2) (multiplied by $10^{10}$). (b) Comparison of the total evaporation residue cross sections $\sigma_{\rm {ER} }=\sum \sigma_{xn}$. The squares and circles denote the experimental data of Dubna \cite{Ogan15,CaAm} and the diamonds denote the data of CAFE2 in Lanzhou \cite{Gan26}.      }
 \end{figure} 

We firstly study the systematic behavior the survival probability of DNS for different projectile-target combinations. In Fig. 1, we show the contour plot of the macroscopic part of the survival probability $P_{\rm mac}$. One sees that the survival probability systematically decreases from a value of one for light-asymmetric fusion systems to about zero for heavy nearly-symmetric projectile-target combinations such as $^{138}$Ba+$^{154}$Sm (with a value of about $ 10^{-14}$). The red and the blue circles denote known hot and cold fusion reactions producing SHN with $Z\ge 110$, respectively. The predicted survival probability $P_{\rm mac}$ in hot fusion reaction $^{48}$Ca+$^{237}$Np  is larger than that of $^{70}$Zn+$^{209}$Bi  by a factor of about $2000$ due to larger mass-asymmetry $\eta$ and consequently smaller effective fissility parameter \cite{Bass74,Swiat82}. Very recently, the evaporation residue cross sections of fusion reactions $^{50}$Ti + $^{242}$Pu  and  $^{54}$Cr + $^{238}$U that lead to the same compound nucleus $^{292}$Lv have been measured and it is found that the cross section of $^{50}$Ti + $^{242}$Pu is 
approximately 15 times higher than that of $^{54}$Cr + $^{238}$U at close excitation energy \cite{Ogan25}. The corresponding $P_{\rm mac}$ of $^{50}$Ti + $^{242}$Pu is about 5 times higher than that of $^{54}$Cr + $^{238}$U, which indicates that the systematic behavior of $P_{\rm mac}$ proposed in this work is reasonable. For the three reactions leading to the synthesis of element 119 shown in Fig. 1, the $P_{\rm mac}$ of $^{45}$Sc+$^{249}$Cf is larger than those of $^{50}$Ti+$^{249}$Bk and $^{54}$Cr+$^{243}$Am by a factor of 7 and 43, respectively.

In Fig. 2, we show the predicted capture excitation function and the corresponding evaporation residue cross sections  with EBD3 for reaction $^{48}$Ca+$^{243}$Am. The arrow in (a) denotes the capture barrier height $V_B=196.85$ MeV from EBD3  which is very close to the corresponding results from BW91 \cite{BW91} and Bass potential \cite{Bass80}. When the incident energy is lower than $V_B$ the capture cross sections fall exponentially with decreasing of $E_{\rm c.m.}$ due to quantum tunneling  effects. The circles show the energy dependence of the predicted production probabilities. At energies well above the capture barrier, the production probability decreases with incident energy due to the decreasing survival probability of the compound nucleus at higher excitation energy according to Eq.(6). At sub-barrier energies, the structure factor $Y$ decreases with decreasing of incident energy according to Eq.(7). From Fig. 2(b), one sees that the measured total evaporation residue cross sections $\sigma_{\rm {ER} }=\sum \limits_x  \sigma_{xn}$ are remarkably well reproduced by EBD3. Here, $\sigma_{xn}$ denotes the cross section of respective neutron-evaporation channel.   
 
\begin{figure}
	\setlength{\abovecaptionskip}{0cm}
\includegraphics[angle=0,width=\textwidth]{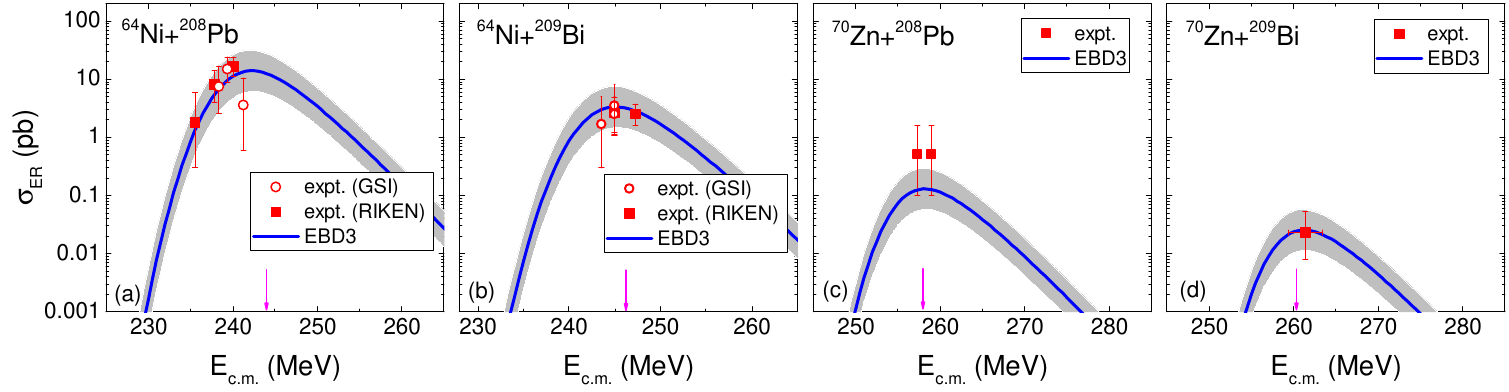}	
 
	\caption{ evaporation residue cross sections for cold fusion reactions  $^{64}$Ni + $^{208}$Pb \cite{Hoff98,Mori04},  $^{64}$Ni + $^{209}$Bi \cite{Hoff02,Mori04},  $^{70}$Zn + $^{208}$Pb \cite{Hoff02}, and  $^{70}$Zn + $^{209}$Bi \cite{Mori09}. The scattered symbols denote the experimental data. The curves denote the predictions of EBD3 and the shadows denote the uncertainties. The arrows denote the positions at $E_{\rm c.m.}=V_B+\Delta$. }
\end{figure}

 \begin{figure}
	\setlength{\abovecaptionskip}{0cm}
\includegraphics[angle=0,width=1\textwidth]{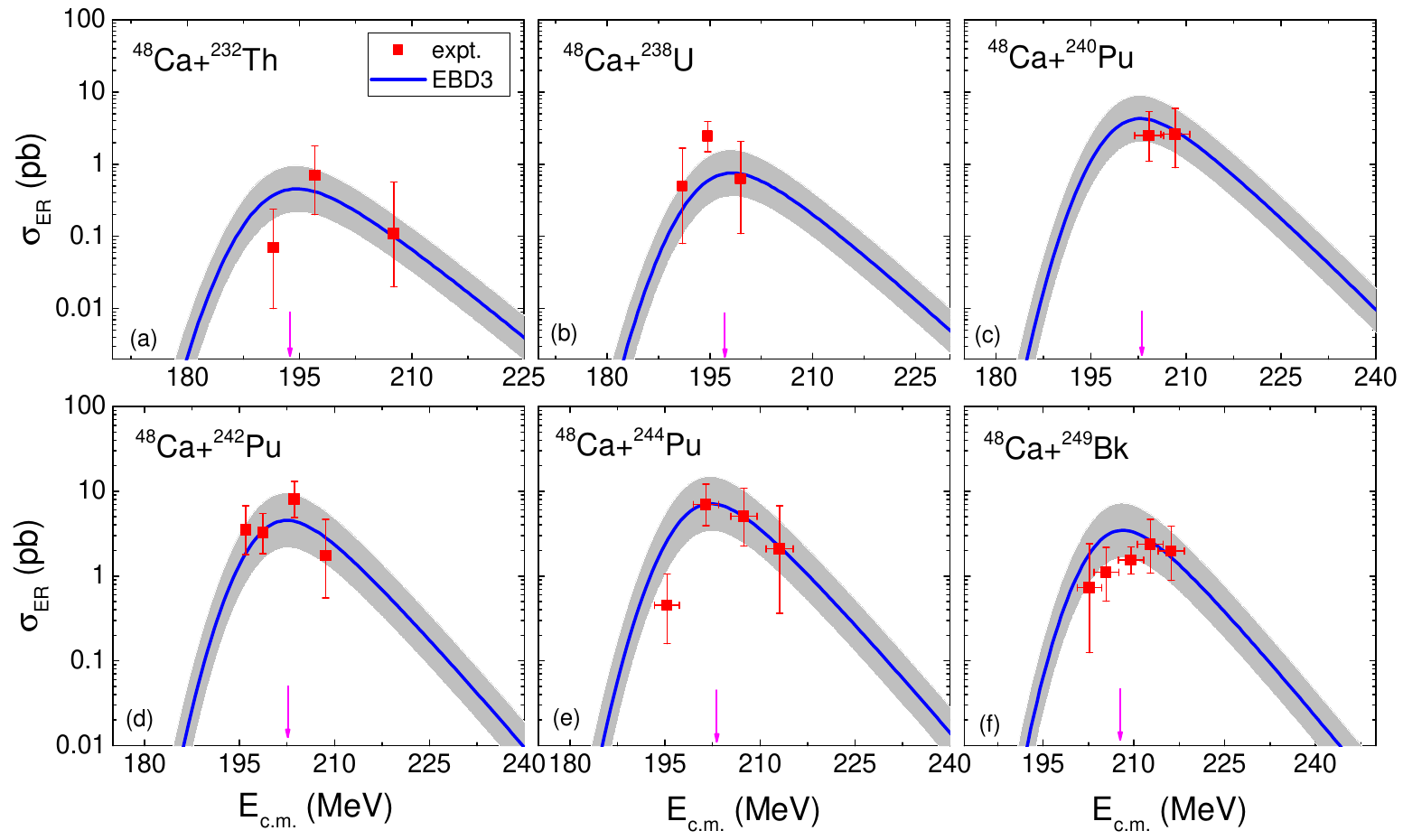}	
 	\caption{  Total evaporation residue cross sections for hot fusion reactions $^{48}$Ca + $^{232}$Th \cite{CaTh}, $^{48}$Ca + $^{238}$U \cite{CaU}, $^{48}$Ca + $^{240,242,244}$Pu \cite{CaU,CaPu240,Ogan15}, and $^{48}$Ca + $^{249}$Bk \cite{Ogan15}. The squares and the curves denote the experimental data and the predictions of EBD3, respectively. The arrows denote the positions at $E_{\rm c.m.}=V_B+B_{\rm f}+\Delta$. }
 \end{figure}   
 
 \begin{figure}
 	\setlength{\abovecaptionskip}{0cm}
 	\includegraphics[angle=0,width=1\textwidth]{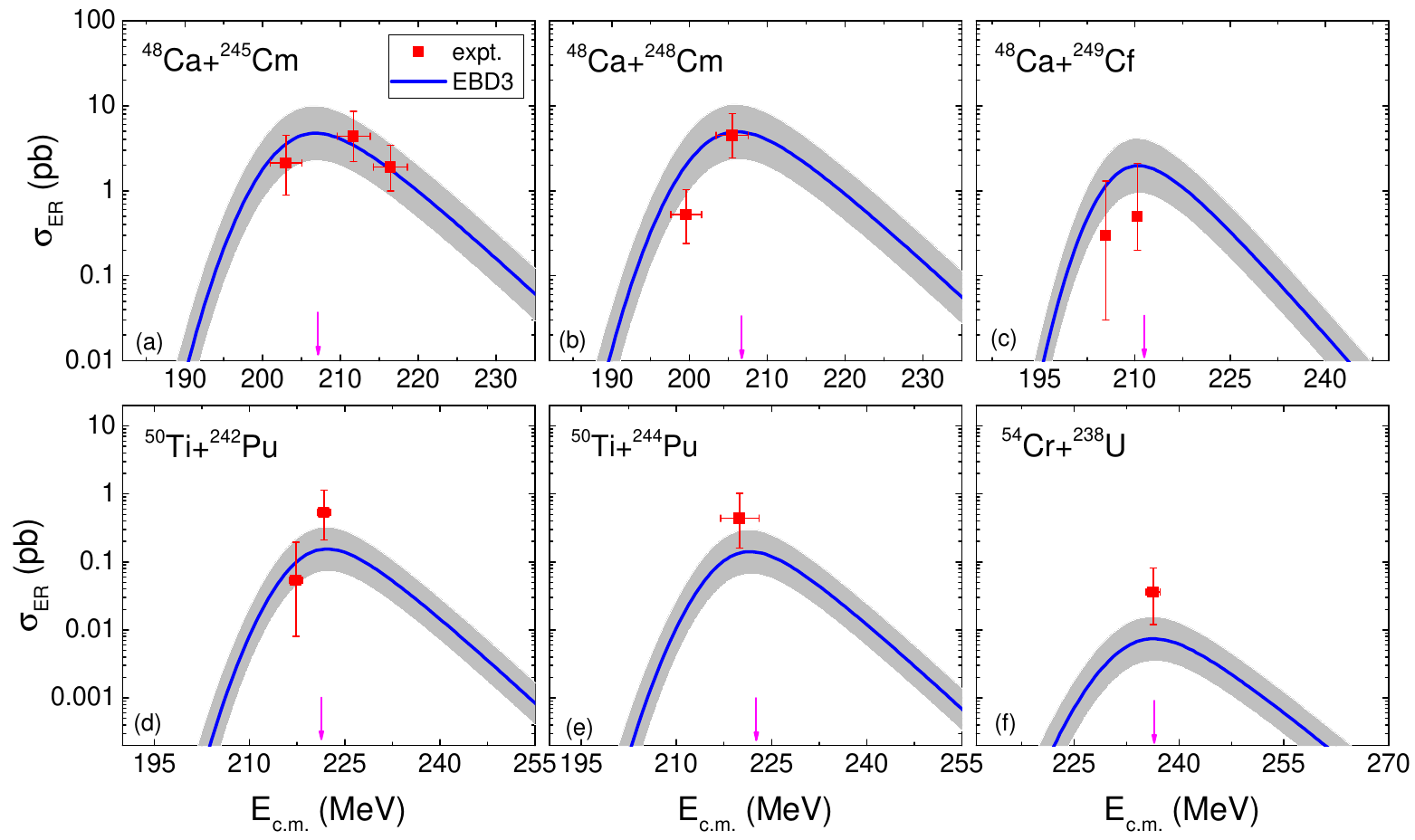}	
 	\caption{ The same as Fig. 4, but for  $^{48}$Ca + $^{245,248}$Cm \cite{CaCm,Ogan15}, $^{48}$Ca + $^{249}$Cf \cite{Ogan15}, $^{50}$Ti + $^{242,244}$Pu \cite{Ogan25,Ogan26,TiPu}, and  $^{54}$Cr + $^{238}$U \cite{Ogan25}.  }
 \end{figure}
    
 In Fig. 3, we show the predicted  $\sigma_{\rm {ER} }$ for cold fusion reactions  $^{64}$Ni + $^{208}$Pb,  $^{64}$Ni + $^{209}$Bi,  $^{70}$Zn + $^{208}$Pb, and  $^{70}$Zn + $^{209}$Bi. The measured maximum cross sections falls from $16.5_{-5.5}^{+7.3}$ pb of $^{64}$Ni + $^{208}$Pb to about 23 fb of $^{70}$Zn + $^{209}$Bi with increasing of charge number $Z$. The solid curves with shadows denote the predictions of EBD3 according to Eq.(1), which are in good agreement with the experimental data. The value of $P_{\rm mac}$ for $^{64}$Ni + $^{208}$Pb is 19.4 times higher than that of  $^{70}$Zn + $^{209}$Bi, simultaneously, both $B_{\rm f}$ and $B_{\rm cap}$ of the former system are higher than those of the latter by more than 1 MeV. It results in that the production probability $P$ of the former is 369 times higher than that of the latter at the incident energy of $E_{\rm c.m.}=V_B$. In addition, the capture cross section $\sigma_{\rm {cap} }$ of the former is  1.7 times larger than that of the latter at $E_{\rm c.m.}=V_B$, due to the relatively larger average barrier radius $R_B$ \cite{EBD2} which is related to the Bass parameter. The arrows denote the positions at $E_{\rm c.m.}=V_B+\Delta$. The capture barrier heights $V_B$ and shell gaps $\Delta$ in the compound nuclei are listed in Table I. One notes that the positions of arrows are close to the corresponding peaks of $\sigma_{\rm {ER} }$. From Table I, we note that the predicted shell gaps $\Delta$ are smaller than 3.5 MeV for all 24 reactions under consideration. It means that the optimal incident energies for cold fusions are slightly higher than the corresponding capture barrier heights $V_B$. Based on the macroscopic-microscopic calculations, it is also found that the shell effects in the entrance channel result in fusion-barrier energies at the touching point that are only a few MeV higher than the ground state for compound systems near $Z = 110$ in cold fusion reactions, and no significant “extra-extra push” energy is needed to bring the system inside the fission saddle point \cite{Moll97}.

In Fig. 4 and Fig. 5, we show the predicted total $\sigma_{\rm {ER} }$ for 12 hot fusion reactions. The experimental data can also be well reproduced by EBD3, which indicates that the proposed empirical production probability in Eq.(3) is reasonable for these fusion reactions. The arrows in Fig. 4 and Fig. 5 denote the positions at $E_{\rm c.m.}=V_B+B_{\rm f}+\Delta$. One sees that the positions of these arrows are close to the corresponding peaks of $\sigma_{\rm {ER} }$, which implies that the extra-push is required to form compound nuclei in hot fusion reactions. In Table I, we also list the predicted optimal excitation energies $E^*_{\rm opt}$ for 24 fusion reactions. Here, we define an effective fusion barrier height as $V_{\rm eff}=V_B+\delta B_{\rm f}+\Delta$ with $\delta=0$ for cold fusion and 1 for hot fusion, which represents the total obstacle that must be overcome throughout the entire fusion process from infinity between two nuclei to the formation of a compound nucleus. We note that the root-mean-square deviation (RMSD) between $V_{\rm eff}$ and $E^*_{\rm opt}-Q$ is only 0.76 MeV for all of 24 reactions.
This consistency check confirms that $V_{\rm eff}$ captures the dominant energy scale governing the optimal incident energy within the EBD3 framework, suggesting it as a convenient estimator for unmeasured systems. For the reactions $^{48}$Ca + $^{244}$Pu, $^{48}$Ca + $^{245,248}$Cm and $^{48}$Ca + $^{249}$Bk, the fission barrier heights $B_{\rm f}$ are comparable with the depths of capture pocket $B_{\rm cap}$ and therefore the scale factor $s \approx 1$. For $^{50}$Ti + $^{242}$Pu and  $^{54}$Cr + $^{238}$U, the values of $B_{\rm cap}$  are obviously smaller than $B_{\rm f}$, particularly for $^{54}$Cr + $^{238}$U with $B_{\rm cap}=3.80$ MeV, which results in significantly decreasing of the scale factor $s$.  The value of $s$ is 0.29 for $^{50}$Ti + $^{242}$Pu and $s=0.12$ for  $^{54}$Cr + $^{238}$U according to Eq.(9).

The fission barrier height $B_{\rm f}$ is a sensitive input parameter in the calculations. One MeV variation of $B_{\rm f}$ can result in the cross sections change by about one order of magnitude. We note that the uncertainties of fission barrier heights from different models significantly increase for extremely neutron-deficient nuclei due to isospin effects partly. For example, the calculated $B_{\rm f}$ with FRLDM \cite{Moll15} for $^{218}$U is higher than the prediction of WS4 combining an empirical correction \cite{Wang24} by 4.37 MeV, while the results of the two models are comparable for $^{238}$U. Considering the uncertainty of $B_{\rm f}$ due to isospin effects, we take the minimum $B_{\rm f}^{\rm min}$ among the WS4 predictions for the compound nucleus  $(N, Z)$ and its four isotonic neighbors $(N \pm 1, Z)$ and $(N \pm 2, Z)$ as the value of $B_{\rm f}$ in the calculations, i.e. $B_{\rm f}=B_{\rm f}^{\rm min}$  for extremely neutron-deficient nuclei with $N-Z < 60$. Here, $N$ and $Z$ denote the neutron and proton numbers of the compound nucleus, respectively. The predicted $\sigma_{\rm {ER} }$ for cold fusion reactions and $^{40}$Ar-induced systems can be significantly improved by using $B_{\rm f}^{\rm min}$ in the calculations. In addition, the shell damping effect in fission barrier is taken into account for hot fusion reactions with high excitation energies. If the excitation energy $E_B^*$ of the compound nuclei at   $E_{\rm c.m.}=V_{\rm eff}$ is larger than $2E_D$, the fission barrier is reduced according to $B_{\rm f}=B_{\rm f}^{\rm WS4}\left( 2E_D/E_B^*\right)^2/x$ for SHN, where the quadratic scaling follows from the expectation that the shell corrections are damped at high excitation energy \cite{WangB25} and $x$ denotes the fissionability parameter \cite{Wang24}.

\begin{figure}
	\setlength{\abovecaptionskip}{0cm}
\includegraphics[angle=0,width=1\textwidth]{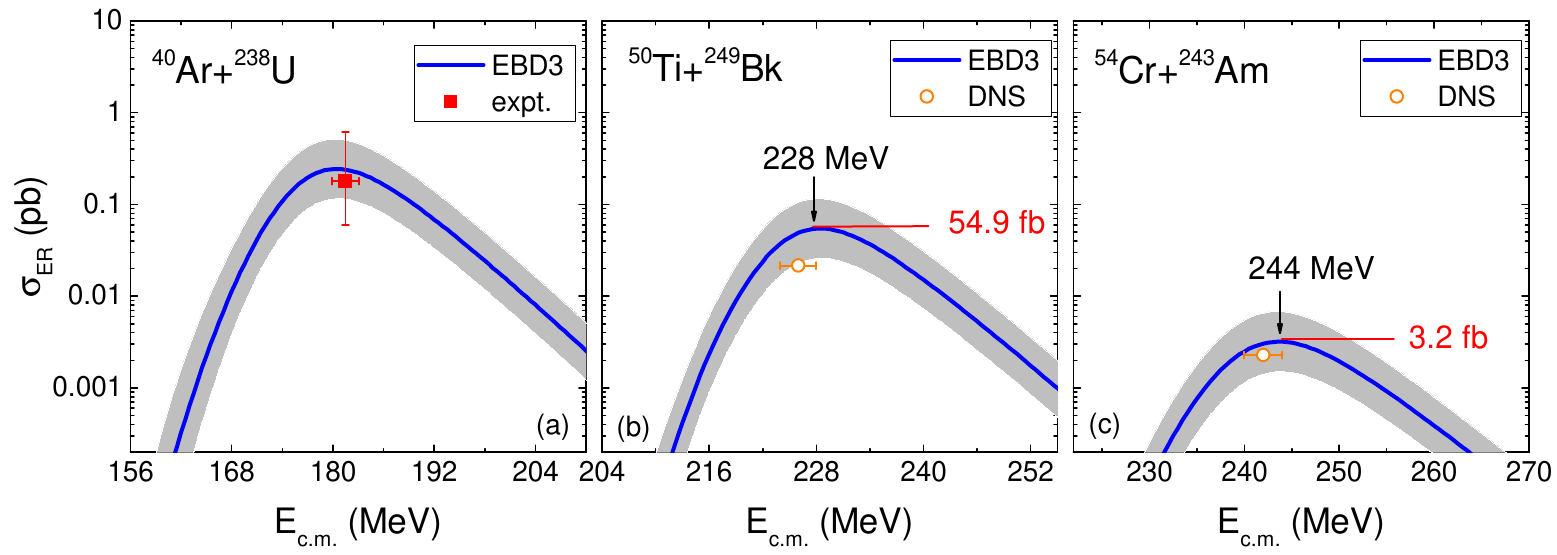}	
	\caption{ Predicted total evaporation residue cross sections for reactions  $^{40}$Ar + $^{238}$U \cite{Ogan24},  $^{50}$Ti + $^{249}$Bk and  $^{54}$Cr + $^{243}$Am. The circles denote the predictions of DNS model \cite{ZhangHF26} . }
\end{figure}

In Fig. 6, we show the predicted  $\sigma_{\rm {ER} }$ for $^{40}$Ar + $^{238}$U,  $^{50}$Ti + $^{249}$Bk, and  $^{54}$Cr + $^{243}$Am. The square in Fig. 6(a) denotes the measured evaporation residue cross section of $^{40}$Ar + $^{238}$U which can be reproduced remarkably well by EBD3. One sees that the measured $\sigma_{\rm {ER} }$ of $^{40}$Ar + $^{238}$U is smaller than the maximum cross section of $^{64}$Ni + $^{208}$Pb by two orders of magnitude, with $E^*_{\rm opt}$  higher than the latter by 30 MeV and $B_{\rm f}$ lower than the latter by 2 MeV, although the mass asymmetry is larger. The optimal incident energies to form SHN with $Z=119$ are 228 MeV and 244 MeV for $^{50}$Ti + $^{249}$Bk and  $^{54}$Cr + $^{243}$Am, respectively. With a value of 3.2 fb, the predicted $\sigma^{\rm max}_{\rm ER}$ of $^{54}$Cr + $^{243}$Am is about 17 times lower than that of $^{50}$Ti + $^{249}$Bk, which is generally consistent with the measured ratio between  $^{54}$Cr + $^{238}$U and $^{50}$Ti + $^{242}$Pu. We note that the cross sections of these two reactions were calculated very recently using the DNS model \cite{ZhangHF26}. The results  are also presented in the figure (denoted by circles) for comparison. One can see that both the maximum cross sections and the optimal incident energies are slightly smaller than the results of EBD3. Given that EBD3 is a phenomenological construction and the DNS model \cite{ZhangHF26} is based on dynamical equations of motion, the fact that the two independent approaches yield cross-section predictions within a factor of about $2-3$ for the unmeasured element-119 reactions lends credence to the general reliability of the estimated production rates.

\begin{table} [!htbp]   	
	\centering	
	\caption{ Barrier heights in EBD3 and the predicted maximum cross sections $\sigma^{\rm max}_{\rm ER}$ for fusion reactions under consideration. $V_B$ and $B_{\rm f}$ denote the barrier height of capture and fission, respectively. $B_{\rm cap}$ denotes the depth of capture pocket in entrance-channel nucleus-nucleus potential. $\Delta$ and $Q$ denotes the shell gap in the compound nucleus and the $Q$-value for fusion, respectively, which are obtained with WS4 \cite{WS4}. $E^*_{\rm opt}$ denotes the optimal excitation energy for producing SHN.  }
	\begin{tabular}{cccccccc}
		\hline\hline
		
		~Reaction~  & $V_B$ (MeV) &$B_{\rm f}$ (MeV)  &  $\Delta$ (MeV)  & $B_{\rm cap}$ (MeV)  & $Q$ (MeV)  & $E^*_{\rm opt}$ (MeV)  & $\sigma^{\rm max}_{\rm ER}$ (pb)  \\
		\hline
		$^{40}$Ar+$^{238}$U  &  173.75   &  3.76   &  1.56   &  7.61   & $-132.48$ &  48.02 &  0.2449 \\
		$^{40}$Ar+$^{243}$Am &  179.14   &  3.40   &  1.86   &  7.04   & $-141.06$ &  44.44 &  0.0598 \\
		$^{48}$Ca+$^{232}$Th &  187.60   &  3.92   &  2.25   &  6.16   & $-157.19$ &  37.31 &  0.4577 \\
		$^{48}$Ca+$^{238}$U  &  191.03   &  4.20   &  2.06   &  5.93   & $-160.78$ &  37.22 &  0.7591 \\
		$^{48}$Ca+$^{237}$Np &  193.45   &  3.94   &  2.12   &  5.70   & $-165.18$ &  35.32 &  0.2691 \\
		$^{48}$Ca+$^{239}$Pu &  195.35   &  4.57   &  2.68   &  5.55   & $-167.42$ &  35.58 &  1.4748 \\
		$^{48}$Ca+$^{240}$Pu &  195.16   &  5.04   &  2.78   &  5.58   & $-166.55$ &  36.45 &  4.3062 \\
		$^{48}$Ca+$^{242}$Pu &  194.79   &  5.08   &  2.81   &  5.66   & $-165.32$ &  37.18 &  4.5436 \\
		$^{48}$Ca+$^{244}$Pu &  194.44   &  5.33   &  3.47   &  5.71   & $-163.90$ &  38.60 &  7.1455 \\
		$^{48}$Ca+$^{243}$Am &  196.85   &  5.24   &  2.51   &  5.50   & $-168.60$ &  35.90 &  5.0094 \\
		$^{48}$Ca+$^{245}$Cm &  198.74   &  5.61   &  2.68   &  5.36   & $-171.27$ &  35.73 &  4.7515 \\
		$^{48}$Ca+$^{248}$Cm &  198.22   &  5.37   &  2.23   &  5.42   & $-169.63$ &  36.37 &  4.9547 \\
		$^{48}$Ca+$^{249}$Bk &  200.28   &  5.37   &  2.04   &  5.27   & $-173.27$ &  34.73 &  3.4685 \\
		$^{48}$Ca+$^{249}$Cf &  202.51   &  5.58   &  2.25   &  5.10   & $-177.08$ &  33.42 &  1.9778 \\
		$^{50}$Ti+$^{242}$Pu &  214.45   &  5.33   &  1.74   &  4.58   & $-182.72$ &  39.28 &  0.1549 \\
		$^{50}$Ti+$^{244}$Pu &  213.99   &  5.18   &  2.87   &  4.67   & $-180.52$ &  40.98 &  0.1417 \\
		$^{54}$Cr+$^{238}$U  &  228.69   &  5.28   &  1.74   &  3.80   & $-195.63$ &  40.87 &  0.0073 \\
		$^{45}$Sc+$^{249}$Cf &  214.07   &  4.42   &  2.68   &  4.75   & $-176.44$ &  44.56 &  0.0444 \\
		$^{50}$Ti+$^{249}$Bk &  220.57   &  5.45   &  1.94   &  4.17   & $-192.04$ &  36.46 &  0.0549 \\
		$^{54}$Cr+$^{243}$Am &  235.84   &  5.45   &  2.94   &  3.29   & $-207.20$ &  36.30 &  0.0032 \\
		$^{64}$Ni+$^{208}$Pb &  240.66   &  5.78   &  3.40   &  2.90   & $-224.32$ &  18.18 &  13.944 \\
		$^{64}$Ni+$^{209}$Bi &  243.45   &  5.17   &  2.87   &  2.69   & $-228.13$ &  16.87 &  3.3367 \\
		$^{70}$Zn+$^{208}$Pb &  256.11   &  4.42   &  1.86   &  2.04   & $-243.32$ &  14.68 &  0.1296 \\
		$^{70}$Zn+$^{209}$Bi &  259.11   &  4.27   &  1.14   &  1.87   & $-246.99$ &  14.01 &  0.0254 \\

		\hline\hline
	\end{tabular}
\end{table}  

In addition, we also note that both the predicted maximum cross sections and the optimal incident energies for $^{50}$Ti + $^{249}$Bk by using another version of DNS model \cite{WangB25} are close to the results of EBD3. In \cite{Wang11}, the evaporation residue cross sections of reactions $^{50}$Ti + $^{249}$Bk, $^{50}$Ti + $^{249}$Cf, $^{54}$Cr + $^{248}$Cm and $^{58}$Fe + $^{244}$Pu were predicted with HIVAP code \cite{Reis85} together with an empirical formula for $P_{\rm CN}$. The predicted $\sigma^{\rm max}_{\rm ER}$ are about $35$ fb, $20$ fb and $5$ fb for $^{50}$Ti + $^{249}$Bk, $^{50}$Ti + $^{249}$Cf, and $^{54}$Cr + $^{248}$Cm, respectively, which are comparable with the results of EBD3. However, for $^{58}$Fe + $^{244}$Pu, the result of EBD3 (0.2 fb) is smaller than that of HIVAP by one order of magnitude. In addition, from Table I, one sees that the more asymmetric fusion system $^{45}$Sc + $^{249}$Cf has a maximum cross section of  $\sigma^{\rm max}_{\rm ER}=44.4_{-13.5}^{+50.1}$ fb at an excitation energy of 44.56 MeV, which could be another promising projectile-target combination for synthesizing element 119 considering the comparable cross section and longer half-life of target comparing with $^{50}$Ti + $^{249}$Bk ($\sigma^{\rm max}_{\rm ER}=54.9_{-29.0}^{+61.3}$ fb).    

\newpage
 
\begin{center}
	\textbf{ B. Uncertainties of model predictions and comparison with other models }\\
\end{center}

\begin{figure}
	\setlength{\abovecaptionskip}{0cm}
\includegraphics[angle=0,width=0.7\textwidth]{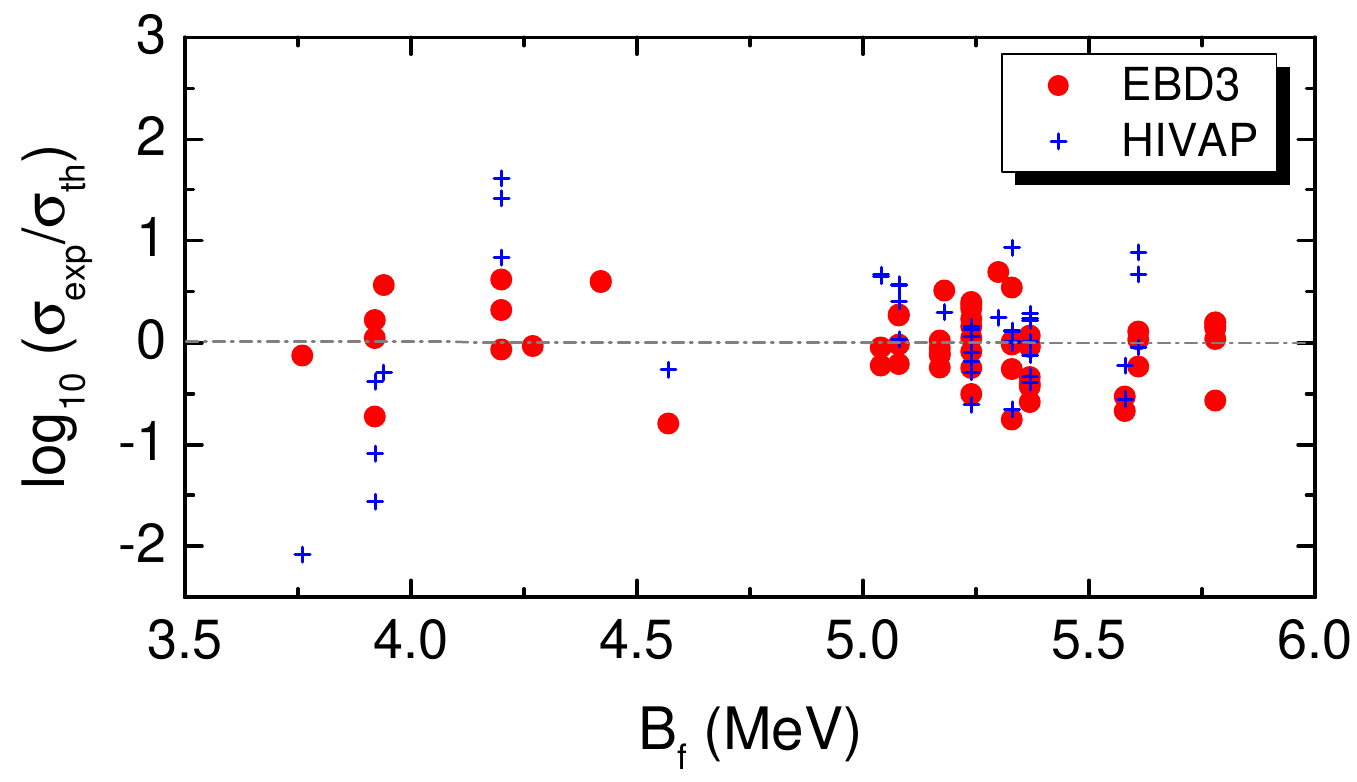}	
	\caption{ Deviations between the predicted evaporation residue cross sections and 58 measured data as a function of fission barrier height. The sold circles denote the results of EBD3. The crosses denote the results of HIVAP for hot fusion reactions. }
\end{figure}

Simultaneously, we investigate the uncertainty of the predicted evaporation residue cross sections with EBD3. On one hand, the uncertainties of  $\sigma_{\rm {ER} }$ can be estimated with the uncertainties of $\sigma_{\rm {cap} }$ and those of $P$, since $\sigma_{\rm {ER} }  =\sigma_{\rm {cap} }\times P$.  In EBD3, the calculations of $\sigma_{\rm {cap} }$ are completely as the same as those in EBD2.2, in which the root-mean-square deviations (RMSD) with respect to 426 datasets of measured capture excitation functions converges to approximately 0.085, corresponding to a relative uncertainty of about $21.6\%$ at energies well above the barrier. The uncertainties of the production probability mainly come from the uncertainties of fission barrier heights. The influence of the uncertainty of $B_{\rm {cap} }$ on $\sigma_{\rm {ER} }$ is relatively small. For example, we note that $\sigma^{\rm max}_{\rm ER}$ will decrease by about $10 \%$ for $^{54}$Cr + $^{243}$Am, if the value of $B_{\rm {cap} }$ is reduced by $10 \%$. In this work, $B_{\rm f}$ is calculated with WS4 based on macroscopic-microscopic framework. For SHN, the macroscopic fission barrier disappears and $B_{\rm f}$ is mainly determined by the ground state shell correction. Considering that one MeV uncertainty in the fission barrier height can lead to a difference of about one order of magnitude in the survival probability and the rms error of WS4 is about 0.3 MeV with respect to known masses, an uncertainty of about $0.3 $ for $P$ could be suitable for estimating the uncertainty of  $\sigma_{\rm {ER} }$ (see the shadows in Fig. 2). On the other hand, the deviations between the predicted cross sections $\sigma_{\rm th}$ and 58 measured data $\sigma_{\rm exp}$ \cite{Fusion} are systematically calculated and shown in Fig. 7. The sold circles denote the results of EBD3 for the 58 measured data. One sees that the deviations for all data are within one order of magnitude and the RMDS is 0.369, which is comparable with the estimation based on the uncertainties of $\sigma_{\rm {cap} } $ and $P$. Here, the corresponding results of HIVAP code \cite{Wang11} for hot fusion reactions are also presented for comparison. We note that for the reactions leading to the synthesis of SHN with relatively higher fission barriers ($B_{\rm f} \geqslant 5 $ MeV), the two approaches yield comparable deviations. At lower $B_{\rm f}$, however, EBD3 systematically outperforms HIVAP. For instance, for $^{40}$Ar + $^{238}$U ($B_{\rm f}$ = 3.76 MeV), the deviation of HIVAP exceeds two orders of magnitude whereas EBD3 stays within a factor of two. The overall RMSD of 0.369 for EBD3 is well within the intrinsic scatter of the experimental data (RMSD $\approx 0.4$), indicating that further refinement within the phenomenological framework would be limited by measurement uncertainties rather than model deficiencies.

\begin{figure}
	\setlength{\abovecaptionskip}{0cm}
\includegraphics[angle=0,width=0.7\textwidth]{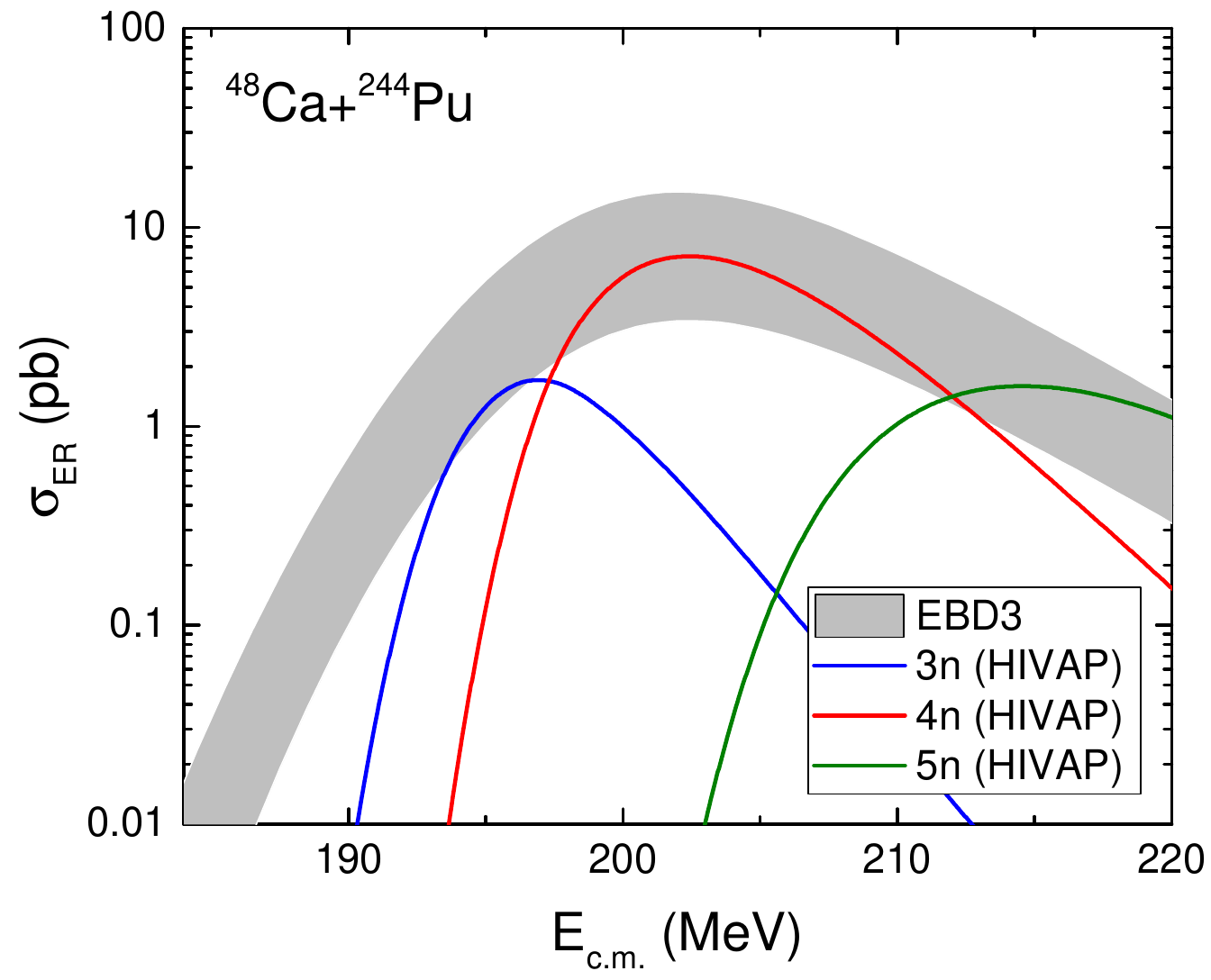}	
	\caption{ Predicted evaporation residue excitation functions for $^{48}$Ca + $^{244}$Pu. The shadows denote the results of EBD3 and the curves denote the results of HIVAP. }
\end{figure}

To see the energy-dependence of $P$ proposed in EBD3, we compare in Fig. 8 the predicted evaporation residue excitation functions from EBD3 and HIVAP for $^{48}$Ca + $^{244}$Pu. For this reaction the scale factor $s=1$. The curves denote the predicted respective neutron-evaporation cross sections from HIVAP and the shadows denote the predictions of EBD3 for total evaporation residue cross sections. One sees that both the predicted maximum cross section and the optimal incident energy from the two approaches are in good agreement with each other, which indicates that the energy-dependence of $P$ in EBD3 is comparable with that in the statistical model.

\begin{figure}
	\setlength{\abovecaptionskip}{ 0 cm}
	\includegraphics[angle=0,width=0.6  \textwidth]{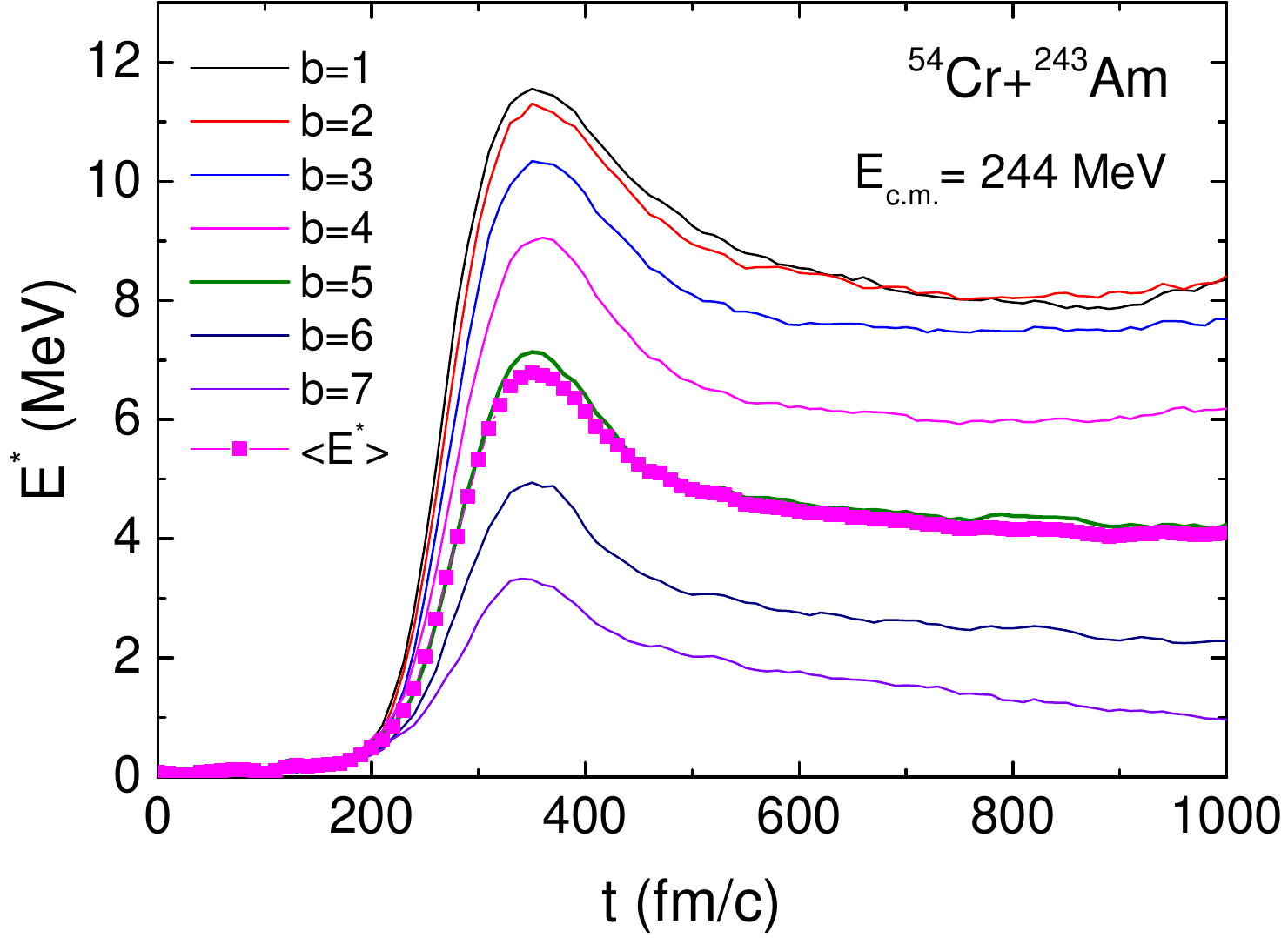}
	\caption{ Predicted excitation energy of the reaction system for $^{54}$Cr + $^{243}$Am at different impact parameter with the ImQMD simulations. }
\end{figure}

In EBD3 calculations, we assume the excitation energy of the DNS with a value of $E_{\rm DNS}^*=20\% E^*$. To verify this hypothesis, we calculate the excitation energy of the reaction system $^{54}$Cr + $^{243}$Am at an incident energy of $E_{\rm c.m.}=244$ MeV by using the ImQMD model \cite{Li20,Yao24}. Fig. 9 shows the time evolution of the excitation energy of the reaction system. The curves denote the results with different impact parameters $b$. Considering the effective barrier radius $R_B = 7.57 $ fm in the calculations of $\sigma_{\rm {cap} }$, we set $b=1-7$ fm in the ImQMD simulations for calculating the excitation energy. One sees that the excitation energy of the reaction system quickly increase around contact between the projectile and target nuclei (at about $t=285$ fm/c), due to the transfer of nucleons. At about $t=350$ fm/c, the excitation energy of the DNS reaches its maximum at a given impact parameter. Then the excitation energy decreases slowly due to breakup of the DNS and particle-evaporation in the de-excitation process. The squares in Fig. 9 denote the average value $\langle E^*\rangle=\frac{ \int 2 \pi b \, E^*(b) \, db}{  \int 2 \pi b\, db}   $ of the predicted excitation energy. The corresponding excitation energy of the compound nuclei at $E_{\rm c.m.}=244$ MeV is $E^*=244-207.2=36.8$ MeV according to the predicted $Q$-value from WS4 \cite{WS4}. We note that the maximum of the average excitation energy is 6.8 MeV, and the corresponding ratio is $18.5\%$ which is close to the assumed value $20\%$. In addition, from the mass-total kinetic energy distributions of fragments in the ImQMD simulations, we note that the quasi-elastic and deep-inelastic scattering plays a dominant role during $t\leqslant 1000$ fm/c. The corresponding capture cross sections from the ImQMD simulations is about  77 mb, which is consistent with the EBD3 prediction of  $\sigma_{\rm {cap} }=64.4 _{-16.7} ^{+22.5}$ mb.

\begin{center}
	\textbf{   C. Limitations of the formula }\\
\end{center}

Despite its satisfactory performance for the 24 reactions studied here, EBD3 has several limitations that should be kept in mind when interpreting its
predictions.

(i) The present formulation employs an average production probability and does not explicitly track the angular momentum dependence of
$P_{\rm CN}$ and $W_{\rm sur}$. This is a tolerable simplification for the cross-section magnitudes discussed here, but may become inadequate for detailed partial-wave analyses.

(ii) The value $E_{\rm DNS}^*=20\% E^*$ has been verified by ImQMD only for the specific system $^{54}$Cr + $^{243}$Am (Fig. 9). Its applicability to other projectile-target combinations with significantly different mass asymmetries or deformations remains to be tested.

(iii) Although $s$ is determined from independently calculated quantities ($B_{\rm f}$, $B_{\rm cap}$, $\Delta$), its functional form is empirical and has been calibrated on systems with $Z \geqslant 110$. Extrapolation to higher $Z$ (e.g., elements 119 and 120) should therefore be regarded as
phenomenologically guided estimates rather than rigid predictions.

(iv) For fusion-fission reactions with $Z < 110$, the relative contributions of quasi-fission and fusion-fission differ from those in the
superheavy regime, requiring re-optimization of the scale factor s before the formula can be reliably applied.

\begin{center}
	\textbf{IV. SUMMARY}
\end{center}
 
 In this work, we have developed a phenomenological formula, EBD3, for the systematic prediction of evaporation residue cross sections in fusion reactions leading to super-heavy nuclei with $Z\ge 110$. The formula adopts the empirical barrier distribution method for capture and introduces a factorized expression for the production probability, which includes macroscopic and energy-dependent components. By incorporating fission barrier height, mass asymmetry, capture pocket depth, and the effective fusion barrier height, EBD3 successfully reproduces 58 experimental data from both hot and cold fusion reactions with high accuracy. The formula also provides a phenomenologically guided estimate of optimal excitation energies and cross section magnitudes, with uncertainties typically within one order of magnitude. From the systematic comparison of the optimal excitation energies, we note that the extra-push energies with values close to the sums of the corresponding fission barrier heights and shell gaps of SHN, are required to form compound nuclei in hot fusion reactions. Predictions for the synthesis of element 119 suggest that highly asymmetric systems such as $^{45}$Sc + $^{249}$Cf and $^{50}$Ti + $^{249}$Bk may offer favorable conditions, with maximum cross sections of about $44$ fb and $ 55$ fb, respectively. For the reactions with heavier projectile nuclei, $^{54}$Cr + $^{243}$Am, the predicted maximum cross sections fall to $3.2$ fb. EBD3 thus serves as a practical and validated tool for guiding future experiments and exploring the limits of nuclear stability. Although with a phenomenological production probability of SHN, the high accuracy and simplicity of EBD3 provide an alternative for further machine-learning analysis. Systematic comparisons with independent microscopic calculations (e.g., TDHF and ImQMD) across a wider range  of systems would help further constrain the empirical components of the formula.

\section*{Data Availability}
The data that support the findings of this article and online calculations with EBD3 are openly
available \cite{Index}.

\begin{center}
	\textbf{ACKNOWLEDGEMENTS}
\end{center}
This work was supported by  Guangxi "Bagui Scholar" Teams for Innovation and Research Project,  Guangxi Natural Science Foundation (Nos. 2026GXNSFBA00640195, 2026GXNSFAA00640236) and National Natural Science Foundation of China (Nos. 12605215, 12465021, 12265006, U1867212). N.W. is grateful to Shan-Gui Zhou, Zhi-Yuan Zhang and Zai-Guo Gan for fruitful discussions.


\begin{thebibliography}{99}

\bibitem{Hoff98} S. Hofmann, Rep. Prog. Phys. \textbf{61}, 639 (1998).

\bibitem{Hoff02} S. Hofmann, F.P. Heßberger, et al., Eur. Phys. J. A \textbf{14}, 147  (2002).

\bibitem{Mori04}  K. Morita, K. Morimoto, et al., Nucl. Phys. A \textbf{734}, 101 (2004).

\bibitem{Mori09} K. Morita, Prog. Part. Nucl. Phys. \textbf{62}, 325 (2009). 

\bibitem{Ogan15}Yu. Ts. Oganessian and V. K. Utyonkov, Rep. Prog. Phys. \textbf{78},  036301 (2015).

\bibitem{Ogan15a} Yu.Ts. Oganessian, V.K. Utyonkov, Nucl. Phys. A \textbf{944}, 62 (2015). 

\bibitem{Ogan17} Yu. Ts. Oganessian, A. Sobiczewski and G. M. Ter-Akopian, Phys. Scr. \textbf{92},  023003 (2017).



\bibitem{Ogan24} Yu. Ts. Oganessian, V. K. Utyonkov, et al., Phys. Rev. C \textbf{109}, 054307 (2024).

\bibitem{Ogan25} Yu. Ts. Oganessian, V. K. Utyonkov, et al., Phys. Rev. C \textbf{112}, 014603 (2025).

\bibitem{Ogan26} Yu. Ts. Oganessian, V. K. Utyonkov, et al., Phys. Rev. C \textbf{113}, 014614 (2026).

\bibitem{Gan26} X. Y. Huang, Z. Y. Zhang, et al., Chin. Phys. Lett. \textbf{43}, 010101 (2026).

\bibitem{Itkis22} M.G. Itkis, G.N. Knyazheva, I.M. Itkisa et al., 
Eur. Phys. J. A \textbf{58},  178 (2022). 

\bibitem{Guolu23} X.-X. Sun and Lu Guo, Phys. Rev. C \textbf{107}, 064609 (2023).

\bibitem{Adam04} G.G. Adamian, N.V. Antonenko, and W. Scheid, 
Phys. Rev. C \textbf{69},  011601(R)  (2004). 

\bibitem{Zag08} V. Zagrebaev and W. Greiner, Phys. Rev. C \textbf{78}, 034610 (2008).

\bibitem{Khu20} J. Khuyagbaatar,  A. Yakushev, et al., Phys. Rev. C \textbf{102}, 064602 (2020).
\bibitem{Tana22} M. Tanaka, P. Brionnet, et al., J. Phys. Soci. Japan \textbf{91}, 084201 (2022).

\bibitem{Wang11} N. Wang,  J.-L. Tian and W. Scheid, Phys. Rev. C \textbf{84}, 061601(R) (2011).

\bibitem{Wangnan12} Nan Wang, E.-G. Zhao, W. Scheid, and S.-G. Zhou, Phys. Rev. C \textbf{85}, 041601(R) (2012).

\bibitem{ZhuL14}L. Zhu,  W.-J. Xie, and F.-S. Zhang, Phys. Rev. C \textbf{89}, 024615 (2014).

\bibitem{Adam20} G. G. Adamian, N. V. Antonenko, et al., Phys. Rev. C \textbf{101}, 034301 (2020).

\bibitem{Nasirov24} A. Nasirov and B. Kayumov, Phys. Rev. C \textbf{109}, 024613 (2024).

\bibitem{WangB25} C.-Y. Zhang, D.-M. Li, et al., Phys. Rev. C \textbf{112}, 014623 (2025).

\bibitem{ZhangHF26} J.-X. Li, F.-Y. Chen, H.-F. Zhang, Chin. Phys. C \textbf{50}, 034102 (2026).  

\bibitem{LiJJ25} J.-J. Li, J.-B. Tian, X.-R. Zhang, et al., Phys. Rev. C \textbf{112}, 064603 (2025).

\bibitem{ZhangFS} M.‑H. Zhang, Y.-H. Zhang, et al., Phys. Rev. C \textbf{109}, 014622 (2024). 

\bibitem{Pei24} D. Guan, J. Pei, Phys. Lett. B \textbf{851}, 138578 (2024). 

\bibitem{ZhangYH26} Y.-H. Zhang, et al., Acta Phys. Sin. \textbf{75}, 020102 (2026). 



\bibitem{Bend99} M. Bender,  K. Rutz,  P.-G. Reinhard, et al., Phys. Rev. C \textbf{60}, 034304 (1999).  
\bibitem{Sam05} M. Samyn, S. Goriely, and J. M. Pearson, Phys. Rev. C \textbf{72}, 044316 (2005).
\bibitem{Pei16} Y. Zhu and J. C. Pei, Phys. Rev. C \textbf{94}, 024329 (2016).

\bibitem{Lala} G.A. Lalazissis, M.M. Sharma, et al.,  Nucl. Phys. A \textbf{608}, 202 (1996). 
\bibitem{Abu12} H. Abusara, A. V. Afanasjev, and P. Ring, Phys. Rev. C \textbf{85}, 024314 (2012).
\bibitem{Lu14}B. N. Lu, J. Zhao, E. G. Zhao, and S. G. Zhou, Phys. Rev. C \textbf{89}, 014323 (2014).
\bibitem{Zhou16} S. G. Zhou, Phys. Scr. \textbf{91}, 063008 (2016).

 
\bibitem{Zhao19} B. Zhao and S.Q. Zhang, Astrophys. J. \textbf{874}, 5 (2019).


\bibitem{Nasirov11} A. K. Nasirov, Mandaglio, et al., Phys. Rev. C \textbf{84}, 044612 (2011).
\bibitem{Loveland} W. Loveland, EPJ Web of Conferences \textbf{131}, 04003 (2016).
\bibitem{Loveland15} W. Loveland, Eur. Phys. J. A \textbf{51}, 120 (2015). 

\bibitem{Moll15} P. M\"oller,  A. J. Sierk, et al., Phys. Rev.  C 91, 024310 (2015).

\bibitem{Wang24} N. Wang, M. Liu, Chin. Phys. C \textbf{48}, 094103 (2024).

\bibitem{Li25} C. Li, X. Luo, T. Li, et al., Phys. Rev.  C \textbf{112}, 034601 (2025).

\bibitem{Li20}  C. Li, J. L. Tian, F.-S. Zhang, Phys. Lett. B \textbf{809}, 135697 (2020).

\bibitem{Yao24} H. Yao,  C. Li, et al., Phys. Rev. C \textbf{109}, 034608 (2024).

\bibitem{Reis85} W. Reisdorf, F. P. Hessberger, et al., Nucl. Phys. A \textbf{444}, 154 (1985).
\bibitem{EBD} H. L\"u, A. Marchix, Y. Abe, D. Boilley, Comp. Phys. Comm. \textbf{200}, 381 (2016).


\bibitem{Bass74} R. Bass, Nucl. Phys. A \textbf{231}, 45 (1974).

\bibitem{Bass80}R. Bass, \textit{Lecture Notes in Physics 117} (Berlin: Springer) pp 281, 1980.

\bibitem{SW04} K. Siwek-Wilczy\ifmmode~\acute{n}\else \'{n}\fi{}ska and  J. Wilczy\ifmmode~\acute{n}\else \'{n}\fi{}ski, 
Phys. Rev. C \textbf{69}, 024611 (2004). 

\bibitem{Wang26} N. Wang, Y.-J. Duan, et al., Chin. Phys. C  \textbf{50}, 044110 (2026). 
 

\bibitem{Pal24} A. Pal, S. Santra, et al., Phys. Rev. C \textbf{110}, 034601 (2024).
\bibitem{Alb20} H. M. Albers, J. Khuyagbaatar, D.J. Hinde, et al., Phys. Lett. B \textbf{808}, 135626 (2020). 





\bibitem{YaoH}H. Yao, H. Yang, and N. Wang, Phys. Rev. C \textbf{110}, 014602 (2024).

\bibitem{Xie26} Y. Xie, N. Wang, Z. Ren, Chin. Phys. C \textbf{50}, 044103 (2026). 

\bibitem{Lest09}J. P. Lestone and S. G. McCalla,  Phys. Rev. C \textbf{79}, 044611 (2009).

\bibitem{WS4}  N. Wang, M. Liu, X. Z. Wu, J. Meng, Phys. Lett. B \textbf{734}, 215  (2014).	 

\bibitem{Swiat82} W. Swiatecki, Nucl. Phys. A \textbf{376}, 275 (1982).

\bibitem{liumin} M. Liu, N. Wang, Z.X. Li et al., 
Nucl. Phys. A \textbf{768}, 80 (2006). 

\bibitem{EP} N. Wang, J.‑M. Chen, M. Liu, Nucl. Sci. Tech. \textbf{36}, 24  (2025).

\bibitem{Mo16} Q. Mo, M. Liu, and N. Wang, Phys. Rev. C \textbf{90}, 024320 (2014).

\bibitem{Rein78} P.-G. Reinhard, Nucl. Phys. A \textbf{306}, 19 (1978).
 
\bibitem{BW91} R.A. Broglia, A. Winther, \textit{Heavy Ion Reactions, Parts I and II, Frontiers in Physics}, vol. 84, Addison–Wesley, 1991. 
 
\bibitem{CaAm} Yu. Ts. Oganessian,  et al.,  Phys. Rev. Lett. \textbf{108}, 022502 (2012). 

\bibitem{EBD2} N. Wang, Chin. Phys. C \textbf{49}, 124106 (2025); arXiv: 2504.13410. 

\bibitem{Moll97} P. M\"oller, J. R. Nix, et al., Z. Phys. A \textbf{359}, 251 (1997).

\bibitem{CaTh} Yu. Ts. Oganessian, et al., Phys. Rev. C \textbf{108}, 024611 (2023).
 
\bibitem{CaU} Yu. Ts. Oganessian, et al., Phys. Rev. C \textbf{70}, 064609 (2004).   

\bibitem{CaPu240}Yu.Ts. Oganessian, et al., Phys. Rev. C \textbf{92}, 034609 (2015).   

\bibitem{CaCm} Yu. Ts. Oganessian, et al., Phys. Rev. C \textbf{69}, 054607 (2004).

\bibitem{TiPu} J. M. Gates, et al., Phys. Rev. Lett. \textbf{133}, 172502 (2024).

\bibitem{Fusion} \url{http://www.imqmd.com/fusion/sigma_SHN2026.html}

\bibitem{Index} \url{http://www.imqmd.com/fusion/}

    \end{thebibliography}
\end{document}